\documentstyle[12pt,epsf]{article}
\renewcommand\thesection{\arabic{section}.}
\renewcommand\thesubsection{\thesection\arabic{subsection}.}
\renewcommand\thesubsubsection{\thesubsection\arabic{subsubsection}.}
\renewcommand\section[1]{\vspace{\topsep}\vspace{\partopsep}
\refstepcounter{section}
{\par  \noindent\normalsize\bf \thesection
\hspace{1em}#1\vspace{\topsep}\par\noindent}}

\newenvironment{refs}
{\vspace{\topsep}\vspace{\partopsep}
{\par \noindent\normalsize\bf  References
\vspace{-\topsep}\par\noindent}
\setlength{\parindent}{-5mm}
\begin{list}{}{\topsep 0pt \partopsep 0pt \itemsep 0pt \leftmargin 5mm
\parsep 0pt \itemindent -5mm}}
{\end{list}}

\renewcommand\subsection[1]{
\refstepcounter{subsection}
{\par \protect\vspace{\topsep}\vspace{\partopsep}
 \noindent\normalsize\bf \it \thesubsection
\hspace{1em}#1\par \noindent}}

\renewcommand\subsubsection[1]{
\refstepcounter{subsubsection}
{\par \protect \vspace{\topsep}\vspace{\partopsep}
\noindent\normalsize \it \thesubsubsection
\hspace{1em}#1\par \noindent}}

\newfont{\sansb}{cmssbx10}
\newfont{\sans}{cmss10}
\newcommand\sun{\odot}

\newcommand\gr{$\gamma$-ray \,}
\newcommand\grs{$\gamma$-rays}
\newcommand\piz{${\pi}^0$ \,}

\begin{document}
\begin{center}
{\large \bf Particle Acceleration and Gamma-Ray Production in Shell 
Remnants\vspace{18pt}\\}
 {H.J.V\"{o}lk\vspace{12pt}\\}
{\sl 
Max-Planck-Institut f\"{u}r Kernphysik, Heidelberg, Germany\\
}
\end{center}

\begin{abstract}

A number of nearby Northern Hemisphere shell-type Supernova Remnants
(SNRs) has been observed in TeV \grs, but none of them could be detected
so far. This failure calls for a critical reevaluation of the theoretical
arguments for efficient particle acceleration and resulting \gr emission
of SNRs which are presumed to be the sources of the Galactic Cosmic Rays. We 
first
discuss diffusive shock acceleration in shell-type SNRs. Observational
upper limits are compared with theoretical predictions for the \gr flux
and found to be roughly consistent. As a next step the empirical arguments
from the observations of X-ray power law continua for Inverse Compton \gr
emission at TeV energies due to electrons are contrasted to the nucleonic
\piz - decay emission from the same objects. Emphasis is given to the
possible problems for VHE \gr production due to the environmental
conditions a SN progenitor finds itself in. Finally, a point is made for
the simplest case of SNe Ia, expected to explode in a uniform
circumstellar medium. In particular the very recently detected Southern
Hemisphere remnant of SN 1006 is compared with Tycho's SNR. On the basis
of the assumed parameters for the two remnants we argue that the TeV
emission from SN 1006 is dominated by Inverse Compton radiation, whereas
Tycho could very well be predominantly a \piz - decay \gr source. 

\end{abstract}

\setlength{\parindent}{1cm}

% Section1

\section{Introduction}
\noindent Shell-type Supernova Remnants (SNRs) have a special significance 
among
expected Galactic \gr sources, because only they have a sufficiently large
{\it hydrodynamical energy output} to replenish the dominant nucleonic
component of the Cosmic Rays (CRs) in the Galaxy. The arguments are as
follows: 
Except possibly for \gr bursts at cosmological distances, SN explosions
are the most violent "events" in the Universe. They have the largest
single release of mechanical energy $E_{SN} \sim 10^{51}$ erg, where a
mass $M_{\rm ej}$ between about one and several $M_{\sun}$ is ejected
with a velocity $V_0 \sim 10^4$ km/sec. Within the Galaxy, assuming a
SN rate $\nu_{\rm SN} \sim 1/30 \, {\rm yr}^{-1}$ this also implies the
largest mean hydrodynamical energy input into the Interstellar Medium (ISM) 
$\langle
\dot{E}_{\rm SN} \rangle \sim 10^{42} {\rm erg \, sec}^{-1} > \langle
\dot{E}_{\rm Stellar Winds} \rangle > \langle \dot{E}^{\rm rot}_{\rm
Pulsars} \rangle $; the factors between each of these energy inputs may
very well reach values of the order of ten. 

\noindent Nevertheless $\langle \dot{E}_{\rm SN} \rangle$ is still only 
marginally sufficient to replenish the CRs against their escape from 
 the Galaxy
$\langle \dot{E}_{\rm CR}\rangle = E_c \cdot V_{\rm conf}/t_{\rm esc} =
E_c \cdot M_{\rm gas}\cdot c/x = 3 \cdot 10^{40}{\rm erg} \,
{\rm sec}^{-1}$
for an energy-independent "grammage" $x=8$\, g cm$^{-2}$, as derived from the
CR composition at GeV energies. In this relation $V_{\rm conf}$ and 
$M_{\rm gas}$ denote the CR confinement volume and the interstellar gas 
mass, respectively. The grammage $x$ decreases with increasing
particle energy as suggested by the observed energy dependence of the CR
secondary to primary ratio $x = 6.9 \cdot (R/{20 \, {\rm GV}})^{-0.6}$ g
cm$^{-2}$ (e.g. Swordy et al., 1990), with $R = E/{Ze}$ denoting rigidity,
as well as by CR propagation theory (Ptuskin et al. 1997).
Therefore the CR {\it source spectrum} is significantly harder than the
steady state particle spectrum observed near the solar system. This 
increases the CR energy input requirement from the
conventionally estimated few percent of $\langle \dot{E}_{\rm SN} \rangle$
(e.g.  Berezinsky et al., 1990) to $\langle \dot{E}_{\rm CR} \rangle \sim
1.5 \cdot 10^{41}$ erg sec$^{-1}$ , i.e. to something between 10 and 20
percent of $\langle \dot{E}_{\rm SN} \rangle$ (Drury et al. 1989). Thus,
energetically there are no plausible sources for the CR nucleon component
in the Galaxy other than SNRs, and their particle acceleration ought to be
very efficient. The same argument does not hold for the CR electrons whose
energy density is lower by 2 orders of magnitude at energies exceeding
$\sim 1$ \, GeV.  Therefore the physical nature of the electron sources is
energetically much less constrained than that of the CR nucleons. 

\noindent Apart from the SNR energetics there are quite strong arguments 
from {\it
acceleration theory}: The energization of nonthermal particles in SNRs is
usually assumed to be due to diffusive shock acceleration at the outer
shock that sweeps up the circumstellar medium during the expansion of the
remnant (for reviews of diffusive shock acceleration theory, see e.g.
Drury, 1983;  Blandford and Eichler, 1987; Berezhko and Krymsky, 1988;
Jones and Ellison, 1991). All models agree that a hard, power law-type
nucleon spectrum with maximum energies around 100 TeV should be produced
during the evolution of a typical SNR in a diffuse Interstellar Medium
(ISM) of not too high density. We shall evaluate the
models, as they have been developped to date, in the next section. 

\noindent Theoretical estimates of the \piz - decay \gr luminosity of SNRs 
(Drury et
al., 1994; Naito and Takahara, 1994) have led to the conclusion that the
resulting \gr flux would be {\it difficult to observe with present
instruments}. We shall discuss the recent TeV searches in the Northern
Hemisphere and compare the model predictions with the upper limits.

\noindent {\it Electron acceleration} in SNRs is more difficult to estimate
quantitatively since the injection efficiency and even the very process of
acceleration for electrons are as yet much less clearly determined.  On
the other hand, it is empirically plausible that electrons are 
copiously accelerated in SNRs. For example, the well-known correlation
between the integrated Far Infrared and radio continuum luminosities of
late type galaxies is dominated by SN precursor stars. The Far Inrared
emission is due to dust reradiation of the stellar UV emission. The radio
continuum is mainly Synchrotron emission by $\sim 10 \, $ GeV diffuse
galactic CR electrons. It is consistent with a hard source spectrum of CR
electrons $\propto E^{-2}$ and a $\sim 10^{-2}$ acceleration efficiency
ratio of electrons to nucleons (Lisenfeld et al., 1996). In addition,
several individual SNRs show X-ray continua which have been attributed to
nonthermal Synchrotron emission by multi-TeV electrons (e.g. Koyama et
al., 1995). The corresponding Inverse Compton (IC) emission might 
therefore be able to swamp the nucleonic \gr emission from SNRs. 

\noindent Altogether, real SNRs are not the ideal spherically 
symmetric
configurations as theoreticians tend to picture them for simplicity. The
circumstellar environment into which a Supernova explodes can in fact be
quite complicated and we shall at least qualitatively attempt to evaluate
the effects of these uncertainties. 

At the Durban ICRC just before this workshop the Cangaroo group has
announced the detection of the Southern Hemisphere SN 1006 in TeV \grs.
This is a tremendously exciting result which suggests that SNRs are indeed
capable to accelerate multi-TeV electrons, as expected from the ASCA
observations in the hard X-ray region. Why should then not multi-TeV
nucleons be accelerated efficiently as well? We may again use the
theoretical estimates for nucleon acceleration, given a set of assumed
parameters for this object that have been derived from observations in
different wavelength regions. Then it appears that the observed TeV \gr
flux could be produced in hadronic interactions only with considerable
difficulties (!). This seemingly paradoxical result, when seen in the
light of the non-detection of other nearby SNRs in the Northern
Hemisphere, can be rationalized if we assume that SN 1006 is an IC \gr
source. This conclusion is of course highly provisional and needs further
careful scrutiny. Tycho's SNR, even though
considerably younger, might have a similar \piz - decay flux as SN 1006. 
If Tycho's X-ray spectrum has only a very small nonthermal component as
suggested by the dominance of emission lines, it might indeed be possible
to detect its \gr emission of hadronic origin in a deep observation with
present instruments like the HEGRA stereoscopic system, or Whipple and
CAT. The results should be known relatively soon. 

% Section 2
\section{Shock Acceleration in Supernova Remnants}
\subsection{Test Particle Models}
\noindent The simplest models treat CRs as test particles in a prescribed,
time-dependent substrate given by a hydrodynamic solution for the
explosion dynamics, for example a Sedov solution. In the first concrete
estimates (Cesarsky and Lagage, 1981; Ginzburg and Ptuskin, 1981; Lagage
and Cesarsky, 1983) it was assumed that the outer SNR shock was locally
plane.  With the pitch angle scattering mean free path $\lambda_{\rm
mfp}(p)=r_{\rm gyro}(p)$ at the
Bohm limit (which implies that a particle is turned around in its motion
parallel to the mean magnetic field after each gyration), an upper limit
to the achievable maximum momentum $p_{max}$ could be calculated by
equating $\tau_{\rm acc}(p_{\rm max}) = {\rm SNR \, age}$.\footnote{In 
reality, the mean free path will somewhat exceed the gyro radius and
therefore $p_{\rm max}$ will be smaller than the above estimated value.
However, a very strong shock should lead to such strong excitation of
scattering MHD waves by the accelerating particles themselves (Bell, 1978;
McKenzie and V\"olk, 1982), that the Bohm limit may indeed approximately
be reached in such a system.} Here
\begin{displaymath}
\tau_{\rm acc}=
\frac{3}{u_{1}-u_{2}}(\frac{\kappa_{1}}{u_{1}}+\frac{\kappa_{2}}{u_{2}})
\end{displaymath}
\noindent , with $\kappa=1/3 \cdot w \cdot \lambda_{\rm mfp} $ denoting the 
diffusion
coefficient in the upstream (1) and downstream (2) region, respectively; 
the particle speed is given by $w$ (e.g. Drury, 1983). For typical values
of the SNR parameters one obtains in this manner $p_{\rm max}\sim 10^{14}
{\rm eV/c}$. 

\noindent The actual {\it nucleon} spectrum in SNRs has first been 
approximately
calculated in the so-called "Onion shell" models (Bogdan and V\"olk,
1983; Moraal and Axford, 1983):  The spectral index $q(t)=3r(t)/(r(t)-1)$
of the accelerated particles at the shock varies with time, following the
time dependence of the shock compression ratio $r(t)$. To simplify the
time dependence of the solution, time is divided into n intervals and
at each time $t_{k} \, (k=0,1,2, ...,n)$ an imaginary shell surface with
radius $R(t_{k})$ is created. The $k^{th}$ shell contains the material
between the shock radii $R(t_{k-1})$ and $R(t_{k})$. After being
accelerated "at time $t_{k}$", the particles in a given radial shell k
remain confined {\it inside} the expanding SNR and therefore loose energy
adiabatically. In this kinematic, quasi-stationary model, the maximum
particle energy depends on the strength of the shock at $t=t_{k}$ and the
actually calculated wave spectrum at this time (V\"olk et al., 1988). To
obtain the spatially integrated spectrum inside the SNR at time t, all
shells k with $t_{k}<t$ are summed up. Particles are finally assumed to be
released from their SNR source when the shock velocity $\dot{R}(t)$
decreases below the Alfv\'{e}n velocity $v_{A}\sim 30-100$ km 
sec$^{-1}$ of the ambient medium. 

\noindent The interesting result was that the final source spectrum 
corresponds approximately to a {\it single power law} in momentum $\propto
p^{-q_{eff}}$ with $4.1 \leq q_{eff} \leq 4.3$ over a large range in particle
momenta above the injection momenta (Fig. 1). At the end of the so-called
sweep-up phase, when the remnant has swept up an amount of interstellar
material comparable to the ejected mass from the explosion, the
spectrum is very hard $\propto p^{-4}$, with the maximum particle energies
achieved. After the end of the Sedov phase, the spectrum is softer, with
the maximum energy $cp_{max}$ only slightly exceeding $10^4$ times the
proton rest energy $mc^2$ for the particular parameters of a Hot ISM. In
such a test particle picture the fraction of injected particles $\delta
\sim 10^{-3}$ has to be restricted to smaller values during early times t
in order to not violate total energy flux conservation since no
backreaction of the accelerated particles is assumed in such models . 

% Fig. 1
\begin{figure}[htbp]  
   \centerline{\epsfxsize=10cm \epsfbox{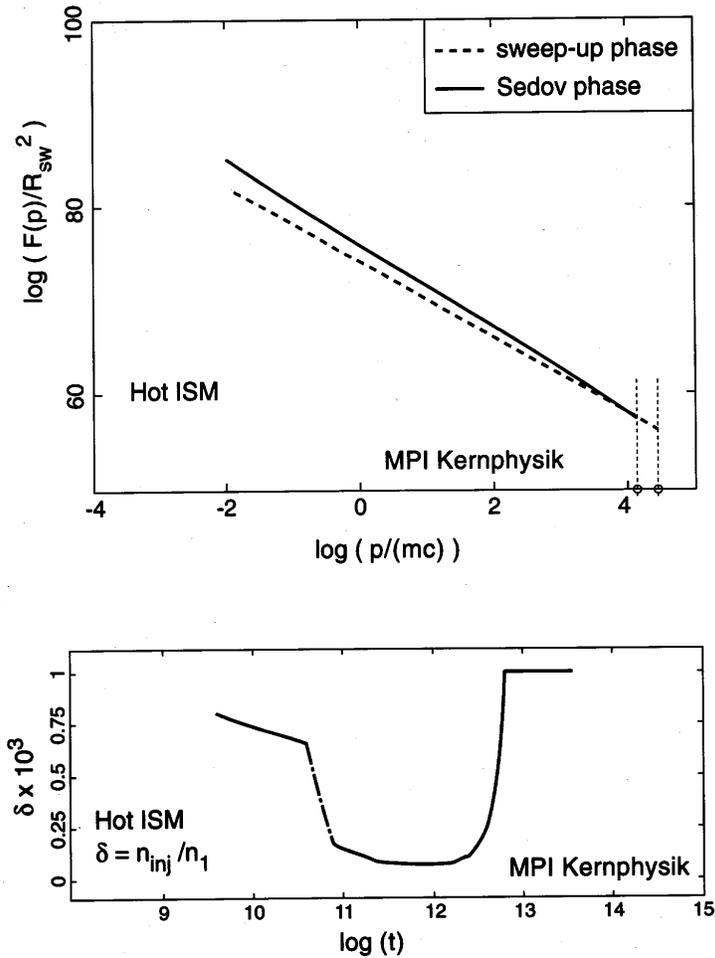}} 
   \caption{Test particle model for the overall (spatially integrated)
proton momentum spectrum $F(p)$ (in arbitrary units) of a SNR in a Hot ISM
as a function of momentum $p$ at the end of the sweep-up phase (dashed
curve) and at release, i.e. at the end of the Sedov phase (solid line). At
sweep-up the spectrum is still quite hard and the upper cutoff higher than
at release, when the spectrum is softer due to continued acceleration of
low energy particles. In order to conserve total energy flux, the
injection rate $\propto \delta$ must be lowered throughout the early times t
of the evolution (adapted from V\"olk et al. 1988). \label{fig1}}
\end{figure}

\noindent For {\it electrons} the determination of $p_{max}$ has to be 
modified, 
since it may be determined by {\it energy loss processes} rather than by the 
system age t alone, as for nucleons. Thus
$\tau_{acc}^{el}(p_{max}) = {\rm Min} \{t, \tau_{loss}(p_{max})\}$
(Reynolds and Chevalier, 1981).

\noindent Reynolds (1996) has applied such models to the X-ray synchrotron 
emission
from SN 1006. He parametrized the scattering mean free path by a factor
${\eta}>1$ times the electron gyro radius, and introduced the speculative
possibility of particle escape from the remnant at some maximum energy.
With these additional parameters in the model he argued that the
observed X-ray continuum emission from SN 1006 could be explained 
by the acceleration of electrons up to 100 TeV in the remnant blast wave.

\noindent A drastic simplification of the Onion Shell model for electrons was
recently made by Mastichiadis (1996): he assumed the accelerated particles
at any given time to be distributed uniformly over the remnant interior,
subject to subsequent adiabatic energy losses. Taking also the magnetic field
strength to be uniform inside the remnant and restricting the injection
rate by globally requiring an electron to proton ratio of $10^{-2}$ at 
release, he could show that the IC yield of
these electrons on the MBR at TeV energies is comparable with the expected
\piz - decay emission from the nucleonic component. 

Even though these electron models simply ignore the  
difficulty of
injecting electrons into the shock acceleration process, they justify
themselves by the fact that multi-TeV electrons seem to be required
empirically to explain the X-ray synchrotron emission from a number of
SNRs (section 5). It seems clear that we have to reckon with
electron IC \gr emission in all SNRs. 

%subsection 2.2
\subsection{Nonlinear 2-fluid Models for Nucleons}
\noindent The test particle models discussed above are not only of a 
kinematic 
and quasi-stationary character, but they also neglect the fact that the 
nucleonic component must have a strong dynamical influence on the SNR 
shock ("backreaction") modifying the acceleration process 
significantly, if this process is to be efficient in the first place. 

\noindent A simple hydrodynamic approximation for the time-dependent 
nonlinear
equations for the {\it coupled dynamics} of the thermal gas (plasma) and
the accelerating CRs is the so-called 2-fluid model (Axford et al., 1977,
1982; Drury and V\"olk, 1981). It consists in taking the kinetic energy
moment of the CR transport equation and introducing the CR pressure
gradient force into the momentum balance for the overall system of plasma
and CRs. Extending this fluid approximation to a 3-fluid model (McKenzie
and V\"olk, 1982), also the scattering wave energy can be
selfconsistently included. Nonlinear shock modification then leads to the
following effects (Fig. 2): In the frame of the shock, the pressure
$p_{c}$ of the accelerating particles has a gradient that decelerates the
incoming flow in the precursor, before the gas pressure, the mass velocity
and the mass density finally jump to their downstream values in a
subshock. Acceleration is the more efficient the more the shock is
smoothed out since in this case irreversible gas heating is reduced and a
higher fraction of the free flow energy must go into CR energy. The total
compression ratio exceeds the canonical value of 4 and therefore at high
energies where the diffusion length is large, the spectrum of accelerated
particles is expected to be be harder than $\propto p^{-4}$, as has been
pointed out by Eichler (1984) and Ellison and Eichler (1984) in their
kinetic discussion of nonlinear shocks. In a SNR this should be the case
during the early evolutionary phases when shock modification is strong. 

%Fig. 2
\begin{figure}[htbp]  
   \centerline{\epsfxsize=10cm \epsfbox{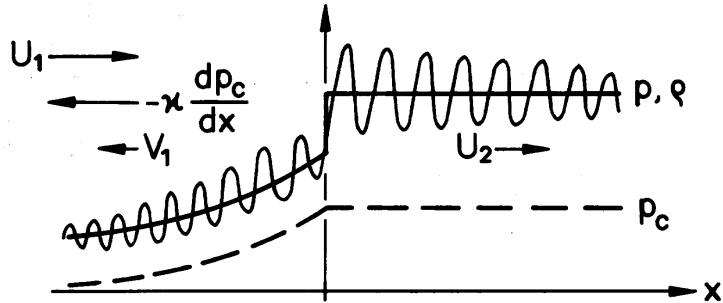}} 
   \caption{Nonlinear effects of accelerated particles as seen in the 
shock frame. The particle pressure gradient $dp_{c}/dx$ produces an upstream
precursor in gas density $\rho$ and pressure $p$. The diffusion current
$\propto -{\kappa}dp_{c}/dx$ excites MHD waves of velocity $V_1$ which are 
swept back and amplified in the subshock at the origin. \label{fig2}} 
\end{figure}

\noindent Particle diffusion into the upstrem medium excites hydromagnetic 
waves
there. They propagate against the incoming flow with the Alfv\'{e}n speed
and get further amplified at the (sub)shock. An important aspect is that
only these selfexcited waves permit efficient acceleration since SNR
shocks have a short life-time, typically of the order of $10^4 {\rm yr}$;
the much weaker average ISM hydromagnetic turbulence alone would make this
process basically ineffective.

\noindent The hydrodynamic approximation has been applied to SNRs in various 
forms
(Drury et al., 1989; Markiewicz et al., 1990; Dorfi, 1990, 1991; Kang and
Jones, 1991). All assume the shock normal to be parallel to the external
magnetic field (parallel shock), in an otherwise spherically symmetric
configuration. This includes adiabatic energy losses in the remnant
interior and diffusive "losses" of particles from the shock into the
interior. Obviously none of these effects can be taken into account in an
application of plane shock theory to such a configuration. Also heating of
the thermal gas by the dissipation of the selfexcited Alfv\'{e}n waves
has been included, even though the diffusion coefficient was maintained at
the Bohm limit. This is no contradiction in the limit of strong wave
excitation. 

\noindent Such model calculations show that for very reasonable injection 
rates
 10-30 percent of the entire hydromagnetic SN energy $E_{\rm SN}$ 
can be
readily converted into CR nuclei during the early evolutionary stages,
when the SNR shock is strong. The ultimate release of CRs into the ISM is
determined by escape and the details of adiabatic cooling during the late
phases where the radiative cooling of the thermal gas and the external
pressure compete in decelerating the remnant expansion.  However, simple
estimates and the numerical results of Dorfi (1991) indicate that an 
efficiency of CR release of about 10 percent is quite possible. 

\noindent Even though the hydrodynamid fluid models constitute a simple 
means to 
calculate the SNR dynamics including CR production, they do not determine
the energy spectra and their temporal evolution. The resulting knowledge of 
the instantaneous shock compression ratio says little about the overall 
spectrum, integrated over the remnant interior, which is relevant for the 
resulting \gr spectra.

\noindent From a quite different angle, such spectra have been estimated by 
Baring
et al. (1997a) using the Monte Carlo approach to particle scattering of 
Ellison and Eichler (1984). This is a plane parallel approximation for
steady shocks and has a kinematic character, assuming the overall dynamics
of the system to be determined in an independent fashion. It is therefore
no surprise that the estimated spectra can only give a qualitative indication
of the true situation of a dynamical system of basically spherical symmetry.
The results were also presented at this workshop by Baring et al.
(1997b). Their emphasis is on \gr cutoffs and the \gr spectral curvature
due to the hardening of the spectrum for successively higher energy
particles. 

%subsection 2.3
\subsection {Dynamical Kinetic Models}
\noindent The first nonlinear and fully time-dependent kinetic 
models for
particle acceleration in SNRs have been presented by Berezhko et al. 
(1994). These authors numerically solved the full combined dynamics of
thermal plasma and CRs, using a phenomenological injection model and the
Bohm limit for the scattering mean free path.  The only limitation of this
calculation is the assumption of a parallel shock everywhere, despite the
restriction to spherical symmetry. Such an approximation clearly cannot
describe asymetries like that of the X-ray continuum emission from SN
1006. Whether a calculation of the full angular dependence might even
imply dynamical instabilities due to the variations in injection rate and
acceleration efficiency with angle is unknown. In the following discussion
we shall ignore this potential difficulty and assume spherical symmetry. 

\noindent The resulting solutions show strong shock modifications with 
increased
compression ratios, and highly efficient nucleon acceleration for moderate
injection rates. The CR spectra at the shock are hardening towards higher
energies. The {\it spatially integrated overall SNR spectrum at late
times} is again an approximate single power law $\propto E^{-q_{eff}}$, but
harder than the corresponding test particle spectra, since now $q_{eff}
\simeq 4.1$. We shall apply this model to the \gr emission of SNRs in the
next section.

%section 3
\section{Nuclear \grs ~ from \piz- decay}
\subsection{Hydrodynamic Approximation}
\noindent The first detailed calculation of the \gr emission from SNRs is 
due to 
Dorfi (1991). He applied the 2-fluid model, adopting the \piz- decay
\gr emissivities appropriate for the observed nucleon spectrum in the ISM 
from Higdon and Lingenfelter (1975), for the 100 MeV energy region. The 
result was that the \gr fluxes from nearby SNRs at $\sim 1$ kpc distance 
were near the lower limit of detectability by the COS-B satellite.

\noindent Using simplified 2-fluid models for the particle acceleration, 
Drury et
al. (1994, referred to as DAV in the following) also calculated the \gr
emission, concentrating on the variations with energy and discussing in
particular the background problems for its detection (see also Naito and
Takahara, 1994). DAV derived the emissivities for various spectral indices
$q_{eff}$, appropriate for the spectra from CR sources, and pointed out
that the diffuse galactic \gr background emission at energies $\gg
m_{\pi}c^2$ decreases with an integral spectral index ${\alpha}\equiv
q-3=1.7$, corresponding to the diffuse galactic CR energy spectrum,
whereas the emission from the sources like SNRs should decrease
considerably slower, with $\alpha$ between 1.1 and 1.3, according to the
test particle results. They concluded that observations of SNRs against
the strong diffuse galactic \gr background should best be done in
the TeV region of currently operating imaging atmospheric Cherenkov
telescopes (IACTs). At energies above 100 MeV, corresponding to satellite
instruments like COS-B or even EGRET, the ${\pi}^0$- decay \gr emission
from SNRs would as a rule be drowned in the diffuse Galactic background
which also would explain the unsuccessful search for SNRs in $ > 100$ MeV
\grs~ so far (Lebrun et al., 1985; Bhat et al., 1985); unfortunately this
is still true today despite repeated efforts in this direction (e.g.
Osborne et al., 1995). 

\noindent It was also pointed out that also in the TeV range a detection 
would be difficult, even
for the best present instruments. Their critical parameter is

\begin{displaymath}
A=\Theta \cdot (\frac{E_{SN}}{10^{51} {\rm erg}}) \cdot(\frac{n}{1 {\rm
cm}^{-3}}) \cdot(\frac{d}{1 {\rm kpc}})^{-2}
\end{displaymath}

\noindent ,where ${\Theta}(t) = E_{c}(t)/E_{\rm SN}$ denotes the fraction of 
total SN energy that is in CRs at any given time during SNR evolution. 

\noindent Since typically $\Theta \sim 0.1$, we have $A\leq 10^{-1}$ or 
possibly
considerably smaller, even for close-by SNRs. Then the flux is indeed
marginal for a typical source that extends over about 1 degree. For $\Theta
E_{\rm SN}=10^{50}{\rm erg}$ and $d=1{\rm kpc}$, the flux of
 TeV \grs~ exceeds $10^{-12} {\rm ph cm}^{-2}{\rm sec}^{-1}$ 
at $n
\geq 0.1 {\rm cm}^{-3}$. The relatively large angular diameter of
SNRs makes them usually {\it extended sources}, making background 
rejection by an IACT difficult. 

%Fig. 3 
\begin{figure}[htbp]  
   \centerline{\epsfxsize=13cm \epsfbox{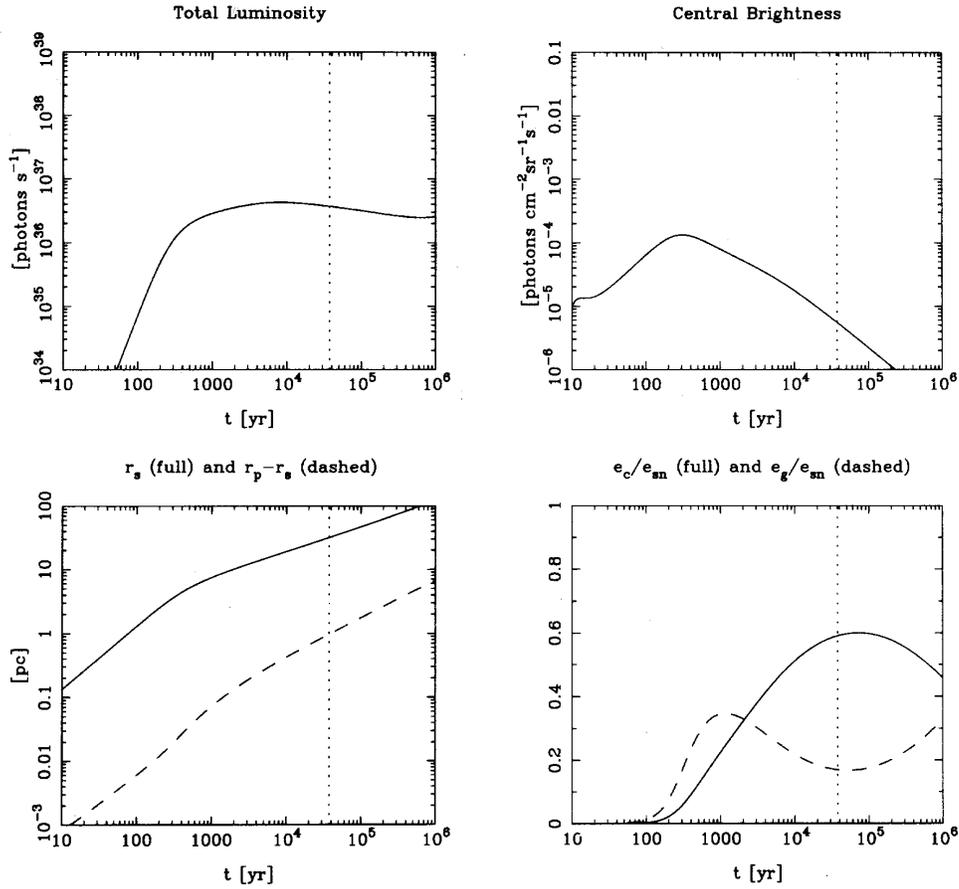}} 
   \caption{Two-fluid results for a SNR. The shock radius $r_s$ increases 
linearly with time t until sweep-up, to slow down to and increase $\propto 
t^{2/5}$ during the Sedov phase that ends whith the onset of 
radiative cooling (dashed vertical line). The total \gr luminosity as 
well as 
the total gas ($e_g$) and CR ($e_c$) internal energies increase sharply 
at sweep-up (cf. Drury et al., 1994). \label{fig3}}
\end{figure}

\noindent In calculating the \gr emissivity the DAV model effectively 
assumes that the
accelerating particles are distributed {\it uniformly} across the SNR for
all energies. Then the \gr energy spectrum is that of a uniform
nucleon energy spectrum and for $E_{\gamma} \gg m_{\pi}c^2$ it is
proportional to the nucleon spectrum. For the spatially integrated \gr
luminosity this concentrates all dynamical effects into the single
efficiency function ${\theta}(t)$ for all \gr energies. In general of
course, the particle diffusion length inside the remnant depends on both E
and t. According to DAV, ${\Theta}(t)$ is very small during the sweep-up
phase, i.e. for very young SNRs with $t<t_0\sim R_0/V_{0}\sim
(\frac{M_{ej}}{{4{\pi}/3}{\rho}_0})/V_{0}$, where ${\rho}_0$ denotes the
external mass density, and $V_{0}$ a characteristic ejecta velocity (Fig.
3). The shock radius $r_s$ still increases linearly with time in this
phase. The \gr flux reaches a very flat maximum during the subsequent
Sedov phase with a slow decrease therafter. 

%subsection 3.2
\subsection{Kinetic Theory}
\noindent The shortcoming of the DAV theory lies in its acceleration aspect. 
Therefore it is important to substitute the 2-fluid model by a kinetic
acceleration model. This has been done by Berezhko and V\"olk (1997a),
based on the theory of Berezhko et al. (1994) described before.  In
addition, the {\it distribution of the SN ejecta velocities} (Chevalier
and Liang, 1989, and refs. therein) was taken into account. In fact, the
fastest ejecta leading the expanding SN material have much higher
velocities than those represented by the mean ejecta speed
$V_0=(2E_{SN}/M_{ej})^{1/2}$. The very high shock velocities which result
from this differential sweep-up at early times $t<t_0$ increase the 
acceleration efficiency
strongly, with a harder overall proton momentum spectrum at $t<t_0$ than
at $t>t_0$ (Fig. 4a), and the \gr flux rises much faster during the
sweep-up phase than in the DAV model at a given \gr energy (Fig. 5). At
$t=t_0$ the flux exceeds that of DAV by a factor of several for the same
parameters $n, E_{SN}, {\rm and} \, d$ due to a harder particle spectrum, an
enhanced CR energy $E_c$, and a larger compression ratio relative to the
hydrodynamic approximation. On the other hand, after a few times $t_0$ the
kinetic \gr flux falls off roughly linearly with time, by a factor of
about 10 over a time period 10 times larger than the time of maximum
during the subsequent Sedov phase. 

%Fig. 4
\begin{figure}[htbp]
   \begin{center}
   \begin{tabular}{cc}
   \epsfxsize=5.5cm \epsfbox{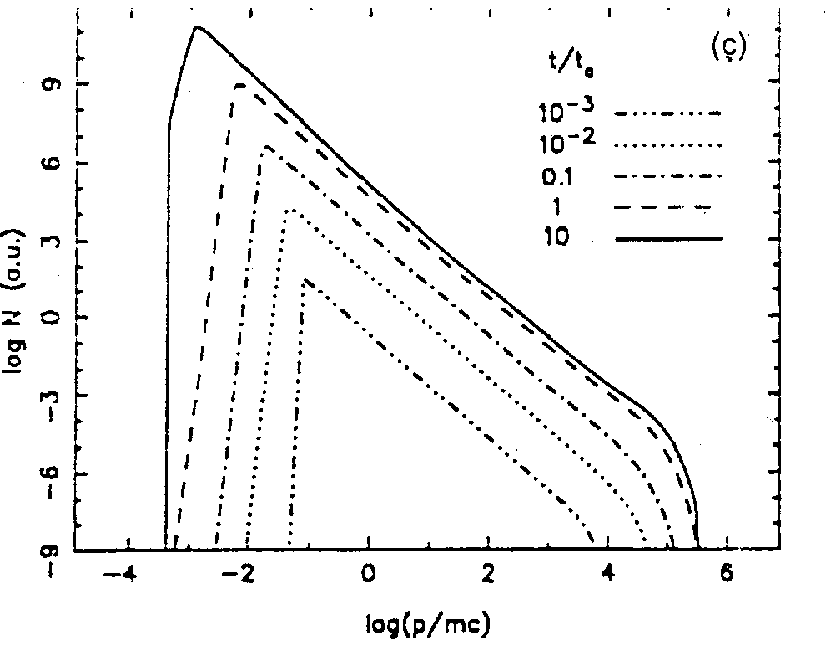} 
&
   \epsfxsize=5.5cm \epsfbox{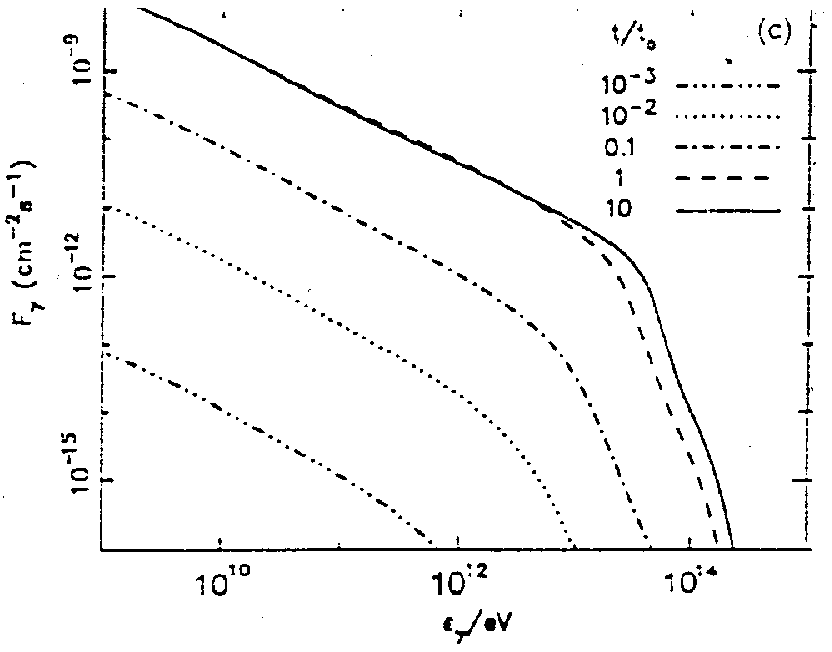} 
   \end{tabular}
   \caption{Overall proton momentum spectrum $N(p)$ versus $p/mc$, for 
various times in units of the sweep-up time $t_0$, in nonlinear kinetic 
theory (left). The resulting integral \piz- decay \gr spectrum 
$F_{\gamma}$  as a function of \gr energy $\epsilon_{\gamma}$ is 
shown on the right (from Berezhko and V\"olk, 1997a). 
\label{fig4}}
   \end{center}
\end{figure}

\noindent Mainly this comes from an increasing lack of spatial overlap 
between 
the interior CR density with the density increase at the shock for all 
but the highest particle energies. While the \gr observability of a SNR 
is therefore reduced for times $t\gg t_0$, the opposite happens for very 
young SNRs with $t<t_0$. Thus we may be able to detect SNRs approaching 
the Sedov phase like Tycho's SNR in nuclear \grs~ (see below).

%Fig. 5
\begin{figure}[htbp]  
   \centerline{\epsfxsize=10cm \epsfbox{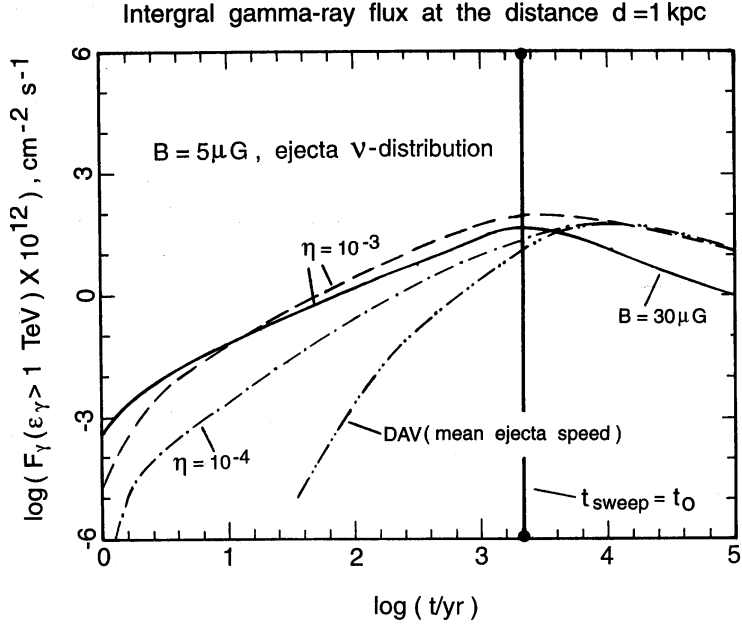}} 
   \caption{Temporal evolution of the $>1 TeV$ \gr flux from a SNR in 
kinetic theory, for various injection rates $\eta$ and two values of the 
external field B, with a distribution of ejecta velocities. This is 
compared to a 2-fluid model with a single mean ejecta speed  
(from Berezhko and V\"olk, 1997a). \label{fig5}} 
\end{figure} 

\noindent The kinetic acceleration theory also allows for the first time the
calculation of the integral \gr spectrum (Fig. 4b). As expected it hardens
with energy similarly to the underlying overall momentum spectrum in the TeV 
range. 
The upper cutoff energy rises $\sim$ linearly with time during sweep-up,
and reaches with the Sedov phase a $\sim$ constant maximum value
$\epsilon_{\gamma}^{max} \propto {n}^{-1/3}\simeq E_{CR}^{max}/10 \simeq
10^{13} {\rm eV}$ for $n\simeq 0.3{\rm cm}^{-3}$. This is again assuming
the Bohm limit and no escape. If, more realistically,
$\lambda_{mfp}={\eta}\cdot r_{g}(p)$ with $1\leq\eta\leq10$ (e.g.
Achterberg et al. 1994), then $E_{CR}^{max}$ and
${\epsilon}_{\gamma}^{max}$ scale as ${\eta}^{-1}$, without spectral
changes for ${\epsilon}_{\gamma}\ll{\epsilon}_{\gamma}^{max}$.
%Section 4
\section{Nuclear \grs ~ and the Observations}
\noindent  When comparing theoretical
models for \piz- decay \gr emission with recent observations in the TeV
range, we shall use the A-parameter while taking into account that
the integral spectral index $\alpha$ is no more a parameter of the theory
but in fact quite small, ${\alpha}=1.1$, or even ${\alpha}\leq 1.0$ at $t
\leq t_0$. Using $\alpha=1.1$ for older, and $\alpha=1.0$ for very young
SNRs, together with ${\Theta}=0.1$, and ${\Theta}=0.2$, respectively,
corresponds to a conservative theoretical estimate of the nucleonic \gr
luminosity. 

\noindent We shall consider here the two SNRs $\gamma$-Cygni and IC 443 
which have
been searched for by the Whipple telescope (Lessard et al., 1997), the
Casa MIA (Borione et al., 1995), Cygnus (Allen et al., 1995), and HEGRA
AIROBICC (Prahl et al., 1997) arrays, and, over parts of the past year, by
the HEGRA stereoscopic system of IACTs (He\ss~ et al., 1997). Both sources
had been detected by EGRET (Esposito et al., 1996) and due to their location
in active star formation regions both are probably the result of the
core collapse of massive stars. The data are shown in the upper panel of
Fig. 6. The data from the HEGRA CT system are still to be considered
preliminary. Both SNRs are extended with a diameter of about 1 degree. 

%Fig. 6
\begin{figure}[htbp]  
   \centerline{\epsfxsize=14.5cm \epsfbox{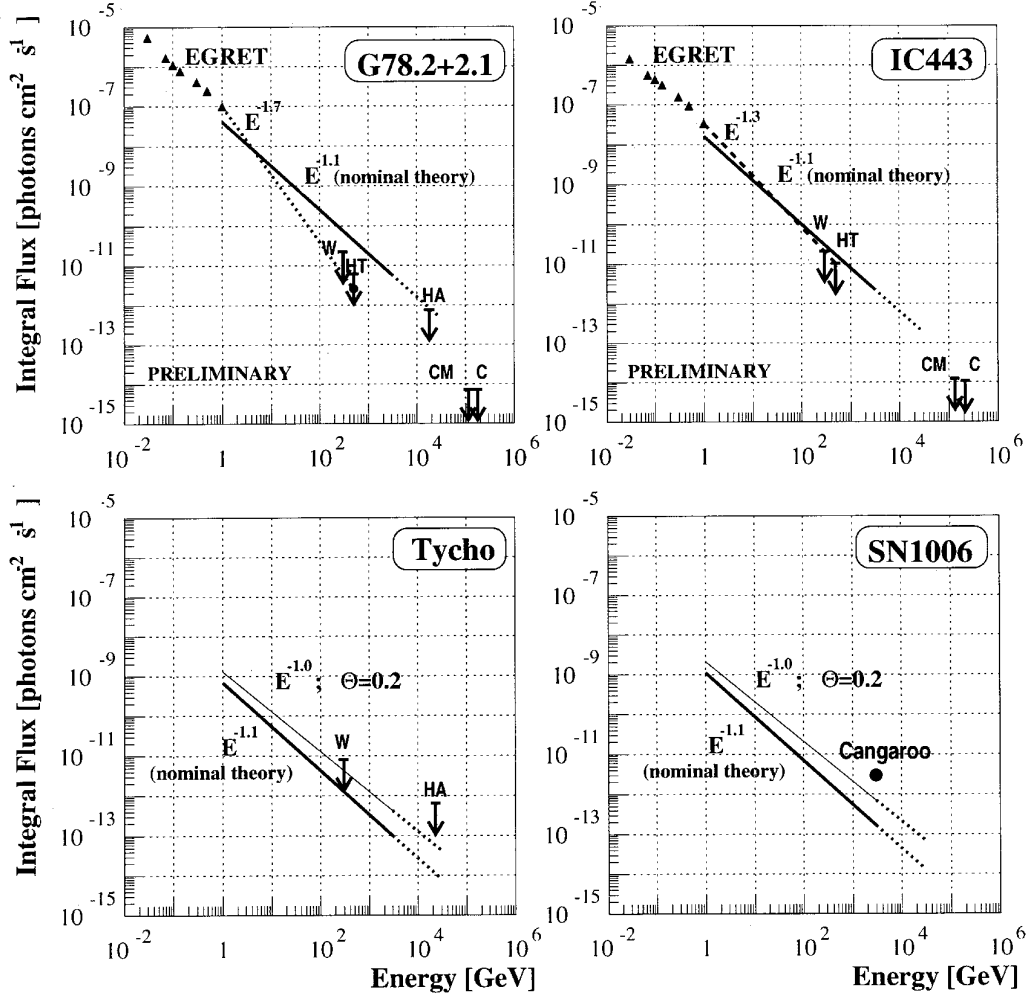}} 
   \caption{Data and theoretical predictions for the \piz - decay 
integral \gr flux from 4 
nearby SNRs. Data points are from EGRET and Cangaroo. Upper limits (ULs) are 
from Whipple, the HEGRA stereoscopic system (HT), HEGRA AIROBICC (HA), 
Casa MIA (CM), and Cygnus (C). The nominal theory corresponds to DAV with 
an integral spectral index of 1.1, and ${\Theta}=0.1$ (thick line). The thin 
line has $\Theta$=0.2 and spectral index 1.0. The point in the UL of HT 
for G78.2+2.1 corresponds to the UL from within the EGRET error circle.
\label{fig6}} 
\end{figure}

\noindent The EGRET source in $\gamma$-Cygni is strongly localized inside 
the SNR,
right at the center of the radio shell, and is geometrically not associated
either with a nearby CO cloud (Cong 1977). A physical interaction of the
SNR with Cong's cloud had been suggested by Pollock (1985). If that was
indeed the case, Aharonian et al. (1994) would predict a sizeable
\gr emission from this cloud both at 500 MeV and at 1 TeV. However,
Claussen et al. (1997) did not find such an interaction from their OH
maser observations at mm wavelengths (as they indeed did for IC 443 and a
number of other SNRs). 

\noindent In Fig. 6 the EGRET data points have been converted to an integral 
spectrum, 
together with the upper limits from the observations at much higher energies 
$> 300$ GeV and a nominal spectrum with parameters given in Table 1.

%Table 1
\begin{table}
\begin{center} 
   \begin{tabular}{|l|l|l|l|l|l|l|} 
      \hline 
      & n [cm$^{-3}$] & E$_{SN}$ [erg] & d [kpc] & $\Theta$ & $\alpha$ &
      A\\\hline
      $\gamma$-Cygni & 5 & $10^{51}$ & 1.5 & 0.1 & 2.1 & 0.2 \\\hline
       IC443& 10 & $2\cdot 10^{50}$ & 1.5 & 0.1 & 2.1 & $9\cdot 10^{-2}$
       \\\hline 
        SN 1006 & 0.4 & $5\cdot 10^{50}$ & 1.8 & 0.1 & 2.1 & $6\cdot 10^{-3}$
       \\\hline 
        Tycho & 1 & $2\cdot 10^{50}$ & 2.3 & 0.1 & 2.1 & $4\cdot 10^{-3} $
       \\\hline 
        Cas A & 30 & 10$^{51}$ & 3.4 & 0.1 & 2.1 & 0.3
       \\\hline
   \end{tabular} 
   \caption{Adopted parameters for the 5 SNRs discussed in the text}
\end{center}
\end{table}

\noindent The 1 GeV EGRET point would connect to the Whipple and HEGRA 
telescope
upper limits by a hypothetical power law with ${\alpha}\sim 1.7$, much too
steep for $\gamma$-Cygni to be considered as an accelerating object for
VHE Galactic CRs. In fact the EGRET source may rather correspond to a
pulsar at this position (see also Brazier et al., 1996). At
${\epsilon}_{\gamma}> 500$ GeV, the HEGRA flux amounts only to about 1/7
of the nominal theoretical flux. This is, unfortunately, still within the
 uncertainties in the A-parameter that are not determined by
acceleration theory. Nevertheless the observed upper limit is
uncomfortably low, to be sure. 

\noindent The case of IC 443 is somewhat different: the spectral index of a 
power
law curve that would connect the HEGRA and Whipple upper limits with the
EGRET 1 GeV point has an index of 1.3, only marginally "too large". Again
it can not be excluded, that the EGRET flux contains the unrelated
emission from a pulsar. In other wavelength ranges IC 443 is a complex
source that may be even consisting of two SNRs (Asaoka and Aschenbach,
1994). Keohane et al. (1996) discovered a very localized hard X-ray
feature using ASCA data, not spatially coincident with the EGRET 95 
percent \gr
detection circle. The SNR interacts strongly with molecular material. 

\noindent Taking the affected molecular gas together with other parameters 
for this
SNR from Fesen (1984) and Mufson et al. (1986), as well as $\theta = 0.1$,
we obtain from Table 1 the nominal \gr spectrum of Fig. 6. 

The Whipple and HEGRA upper limits, at thresholds of 300 and 500 GeV,
respectively, lie by the respective factors of 0.7 and 0.6 below the
nominal theory, easily within all the uncertainties contained in the
A-parameter. We conclude that a deeper observation in the VHE range might
indeed lead to a detection even with present instruments. 
The cases of SN 1006 and Tycho's SNR will be discussed below.
% section 5
\section{Empirical Arguments From X-Ray Power Law Continua}
\noindent Recent observations of hard X-ray continua from several SNRs with 
ASCA (SN
1006: Koyama et al., 1995; IC 443: Keohane et al., 1997; RXJ1713.7-3946:
Koyama et al., 1997), and RXTE and OSSE (Cas A:  Allen et al., 1997; The
et al., 1996, 1997) have probably to be interpreted as nonthermal
synchrotron emission from VHE electrons with energies ranging up to 100
TeV. These electrons should not all come from pulsars, as the example of
SN 1006 shows which is a SN Ia. Thus the energetic electrons should arise
from a different process, most likely diffusive shock acceleration. This
is also consistent with the radio synchrotron emission from entire
galaxies (see Introduction). The difficulty with electrons is that
they cannot be easily injected into the shock acceleration process by
thermal leakage from the hot downstream region, in an analogous manner as
the ions (Levinson, 1994). On the other hand there are many other
instabilities associated with shocks that affect electrons. In particular
lower hybrid waves have been considered (Galeev, 1984; Galeev et al.,
1995, McClements et al., 1997). They might not only pre-energize electrons
to be subsequently, for energies $ > 1$ GeV, accelerated by the robust Fermi
mechanism along with the nucleons, but they might in addition directly
produce nonthermal power laws up to $10^{5}$ GeV, as argued by Galeev
(1984). Ignoring the latter possibility as less likely, if there are $ > 10$
TeV electrons, there should a fortiori exist many more $>10$ TeV nucleons
in such SNRs. Kinematic models for electron acceleration in SNRs have
been constructed by Reynolds and Chevalier (1981), Ammosov et al. (1994),
Reynolds (1996), and Mastichiadis (1996), and have been used to calculate
for example the IC emision from SN 1006 (e.g. Reynolds, 1996; Mastichiadis
and de Jager, 1996) and W44 (de Jager and Mastichiadis, 1997). Indeed a
nonthermal X-ray flux $F_X$ should be accompanied by an IC $\gamma$-ray
flux $F_{\gamma}=F_{X}U_{\rm rad}/U_{B}$ at the corresponding photon
energies ${\epsilon}_{\gamma} \sim h{\nu}_{X}{\nu}_{\rm rad}/{\nu}_{\rm
gyro}$, where $U_{\rm rad}$ and $U_{B}$ are the energy densities of the
radiation field (with characteristic frequency $\nu_{\rm rad}$) and of the
magnetic field, respectively. Time dependent emissions from all
radiative processes, given the respective particle populations, have been
evaluated by Sturner et al. (1997) and Gaisser et al. (1997). 
%section 6
\section {\piz- decay vs. IC \gr fluxes}
\noindent Despite the low proportion of the electronic nonthermal
energy density, the large Thompson cross section and the large number
density of MBR photons can lead to an IC component of the TeV \gr flux
that is quite comparable to that from nucleonic interactions.
Nevertheless, for low $U_{\rm rad}/U_{B}$ the synchrotron channel will
take over most of the electron energy loss. In addition, for high gas
densities the \piz - decay flux will be high. Typically $U_{B}$ will be
enhanced where the density is enhanced. Thus, in principle, for nuclear
\grs~ we should look for SNRs in a high density environment like IC 443 or
Cas A. A straightforward pursuit of this approach may however encounter a
number of difficulties. 
%section 7
\section{Problems}
\noindent We shall here make a short list of possible {\it reductions} of 
the \gr
luminosity relative to the predictions from a blind application of the above 
model. One case, high ambient densities, has already been mentioned above. 

\subsection{Spherical shocks in a uniform B-field}
\noindent  For a basically
spherical SN explosion into a uniform ISM with a homogeneous magnetic
field, the shock normal directions will be in general oblique and - in the
extreme - even be perpendicular ($\perp$) to the external B-field
direction. Thus, first of all, the magnetic compression will vary along
the shock surface. Jokipii (1987) has argued that acceleration may be
faster for effectively $\perp$ shocks for $\eta=\lambda_{\rm
mfp}/r_{g}={\eta}_0>1$, because in this case particle diffusion
across the B-field becomes the dominant process for particle transport
along the shock normal. However it is not clear at all whether for ${\perp}$
shocks scattering waves can also be assumed to be self-excited
effectively, and with the right intensity determined by the required 
parameter
${\eta_0}$. If not, acceleration will be ineffective. Furthermore, for
oblique and ultimately ${\perp}$ shocks, nucleon injection is
increasingly impeded compared to parallel shocks (Bennett and Ellison,
1995; Malkov and V\"olk, 1995). For all these reasons the nonthermal
emission from SNRs will not be spherically symmetric, as clearly visible
in the case of SN 1006. For nucleons it would correspond to a
reduction in overall acceleration efficiency compared to spherically 
symmetric models. For
shock accelerated electrons the X-ray synchrotron and IC \gr lobes of SN
1006 would then occur where the shock is parallel, not where it is
perpendicular, in contrast to the assumptions of e.g. Reynolds (1996)! 
It would be important to measure the polarisation of the X-ray
emission, given these contradictory conclusions. Also the intensity 
distributions should be quite different for the different geometries.

\subsection{Deviation from the Bohm Limit and High Density}
\noindent As discussed before, the upper cutoff energy $E_{CR}^{max} \propto 
1/{\eta}$. 
Thus, for ${\eta}=10$ and $\pi^0$-decay $\gamma$-rays, 
${\epsilon}_{\gamma}^{max}
\simeq$ 2 TeV for an external density $n = 0.3 \, {\rm (H-atom) \, cm}^{-3}$
(Berezhko and V\"olk, 1997a). This will not impede observation at energies
$\leq 0.5$ TeV. However, if in addition the external
number density is increased by a factor $\leq 100$ to $n\leq 30 \, {\rm
(H-atoms) \, cm}^{-3}$, then $E_{max} \leq 0.4$ TeV and \gr detection at 0.5
TeV threshold is no more possible. Thus a lower threshold $\sim 100$ GeV
becomes a necessary condition as for example envisioned in the HESS and 
VERITAS projects. 

%Fig 7.
\begin{figure}[htbp]  
   \begin{center}
   \begin{tabular}{cc}
   \epsfxsize=5.5cm \epsfbox{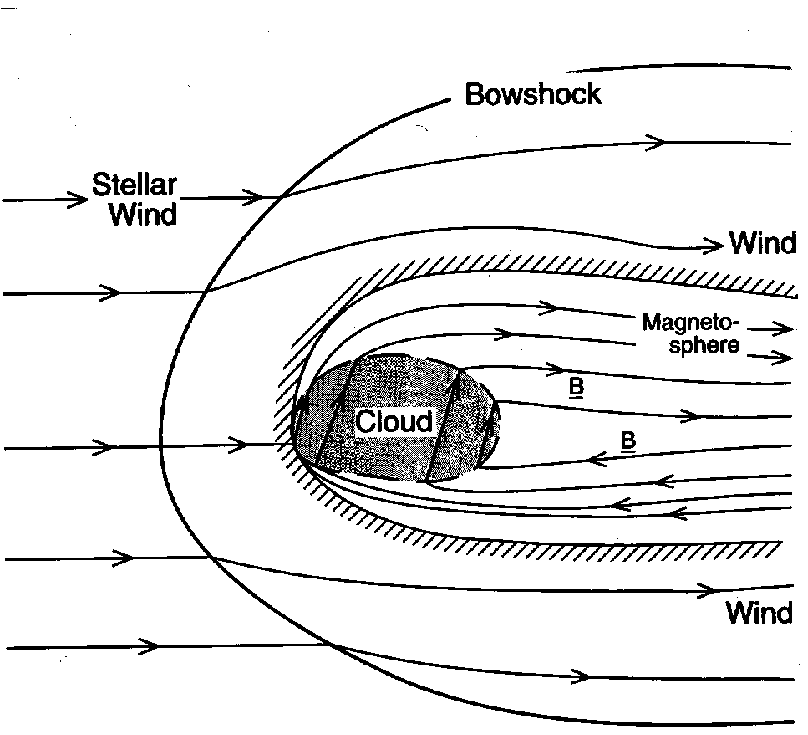} 
&
   \epsfxsize=5.5cm \epsfbox{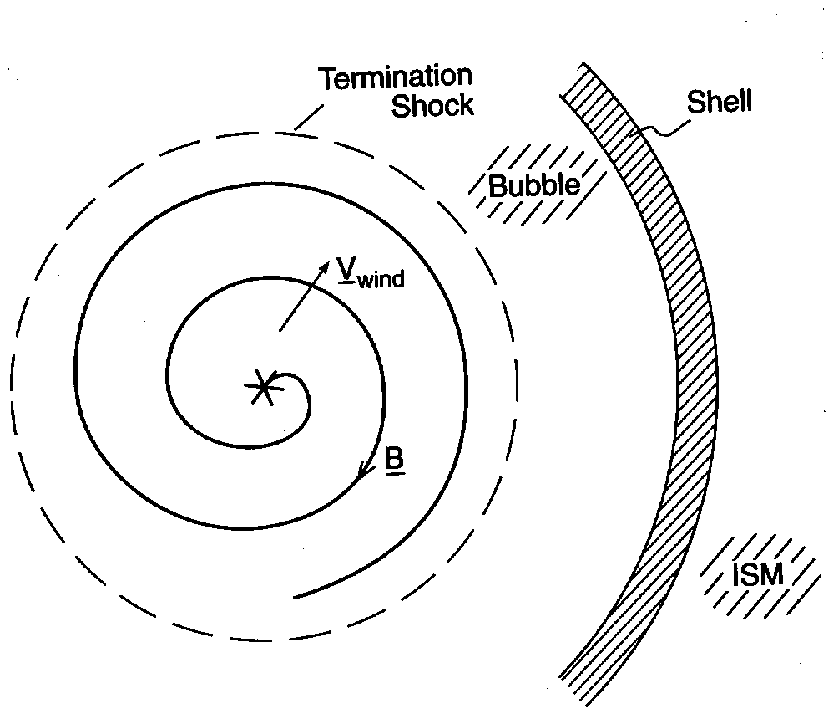} 
   \end{tabular}
   \caption{Flow velocity and magnetic field lines for a supersonic 
stellar wind around a magnetised interstellar cloud (a). Wind bubble 
configuration around a rotating massive star. The wind inside 
the termination shock produces a spiral field configuration. The hot 
shocked wind in the adjacent bubble compresses ISM material into a dense cold
shell seperating the bubble from unperturbed ISM (b). \label{fig7}}
   \end{center}
\end{figure} 

\subsection {Escape}
\noindent In an assumed steady state for a strong shock, the necessary 
existence of
an upper cutoff in the momentum distribution implies the escape of
particles at this upper cutoff. In a time dependent situation that may not
be the case. Nevertheless there might always be a loss of particles at the
upper cutoff due to insufficient wave generation there. Another cause for
particle losses may be given by purely geometrical conditions, for example in
acceleration at a spatially limited shock, like a planetary bow shock in
the solar wind. Finally, in a SNR shock the ionization of the upstream
circumstellar medium may be insufficient if the gas density is high,
despite the ionisation of a radiative precursor by the hot SNR gas. As a
result wave damping due to ion-neutral friction may inhibit wave
excitation for very high energy particles (e.g. V\"olk et al., 1981; 
Draine and McKee, 1993; Drury et al., 1997). Already in moderately dense
clouds ($n\leq10{\rm cm}^{-3}$) this implies a cutoff well below 1 TeV. 
%subsection 7.4
\subsection{Magnetic shielding}
\noindent Gas clouds in the environment of a SNR may be partially 
shielded if their
magnetic field is swept back by the wind from the massive progenitor star.
The SNR will then engulf the cloud but not necessarily produce a great
intensity of energetic particles that could illuminate the cloud along the
magnetic field lines (Fig. 7a).  The shock transmitted into the cloud will
be rather slow and may therefore not produce VHE particles while crossing
the cloud. Thus, even a physically interacting cloud may not necessarily
lead to an enhanced $\pi^0$-decay \gr luminosity to the extent calculated by
Aharonian et al. (1994). 
%subsection 7.5
\subsection{Perpendicular Shock into a Progenitor Wind}
\noindent If the SN progenitor is a massive star of a mass $M\sim 20 
M{\sun}$, it
will strongly modify its environment by its supersonic mass loss. Combined
with the high rotation velocity, the wind will draw out the stellar
magnetic field into an Archimeden spiral geometry (Parker, 1958).  At a
typical distance of $\sim$ 1 pc the radial wind flow will adapt to the
exterior pressure in a termination shock beyond which an expanding hot
bubble of rarefied shocked stellar wind material creates a dense shell of
swept up, cooling interstellar gas (Fig. 7b). The same effect exists for
lower mass SN progenitors, but quantitatively it becomes only important
for massive stars. When the star subsequently explodes as a SN, particle 
acceleration will be
extremely rapid initially (V\"olk and Biermann, 1988) due to the high
B-field $\sim$ 1 Gauss of stellar origin. Possibly this will be true even
at later times (Biermann 1993a, 1993b). Since also the gas density in the
wind is very high initially, $n\propto r^{-2}$, the $\pi^0$- decay \gr
emission may also be very large at early times as pointed out by Kirk et
al. (1995). 

\noindent The SNR evolution in this complex wind/bubble structure has been 
modelled in the kinetic approximation (Berezhko and V\"olk, 1997b), again
assuming acceleration at a parallel shock while maintaining overall
spherical symmetry. Roughly speaking, the \gr luminosity is indeed very
large in the wind during an early episode of a few weeks to months, but
then it decreases to quite a low value in the shocked wind bubble until the
dense shell is reached after $\sim 10^3$ years. Subsequently the high
external gas density leads to a high \gr luminosity which persits until
the shock weakens and the accelerated particles leave the remnant. The
result is a complex time structure of the \gr flux as opposed to the case
of a SNR in a uniform ISM, and the observability of such an object will
depend critically on the phase of its development. Thus non-detection of
such objects in \grs~ at the present time, even though they may look
spectacular in other wavelengths, is not a sufficient argument against
acceleration. We may have looked at it simply at the wrong time! 

\noindent However, this is not the whole story yet. In a wind bubble the 
magnetic
field is largely azimuthal except in the very polar regions, near the
stellar axis of rotation. Therefore not only injection should be strongly
inhibited but possibly also the acceleration itself. The above model
calculations should then only constitute an upper limit to the
true \gr luminosity at any given time. We should perhaps conclude that
massive stars (and their demise as SN) which are amongst the most
impressive fireworks in the sky, are not necessarily the clearest 
fingerprints of
nonthermal processes in SNRs in general. In addition these SN explosions
are likely to leave a pulsar or a black hole behind. Objects like
$\gamma$-Cygni or IC 443 may well be in this category: too complex and
possibly too confused to serve as templates for CR acceleration. 
% section 8
\section{SNIa in a Uniform Medium}
\noindent After this complex digression to uncertainties in the theoretical 
picture and to
astronomical difficulties we should now return to comparatively safe grounds.
These are SNe in a uniform ISM. SN Ia without obvious interaction
with the environment are the natural \gr candidates: old stellar objects of
low mass, disintegrating completely. Two nearby remnants are of special 
interest
here. SN 1006 has been anounced as a detected TeV source by the Cangaroo
group during the Durban conference just before this workshop. Tycho's SNR
may be a similar object yet with different characteristics, even though
only an upper limit exits (Lessard et al., 1997). 
%subsection 8.1
\subsection{SN 1006}
\noindent Let us first discuss the $\pi^0$- decay \gr flux. 

\noindent In order to determine the A-parameter we follow Mastichiadis 
and de Jager
(1996, and references therein) who take $d = 1.8$ kpc, $E_{SN} = 5 \cdot
10^{50}$ erg, and $n = 0.4$ cm$^3$, which gives $A=6.2\cdot 
10^{-3}\cdot
(\Theta /0.1)$, see Table 1\footnote{The catalog of Greene (1996) quotes a
distance range of $1.7<d[{\rm kpc}]<3.1$, a diameter of 30 arc min, and a
radio spectral index of 0.6. The uncertainties in the product $nE_{SN}$
from X-ray data correspond at least to a factor of 2 due to the general
difficulties of modelling the {\it thermal} X-ray emission of SNRs.}. 
With an integral spectral index $\alpha=1.1$ this results in a
theoretically predicted flux of $F_{\gamma,
\pi^0}^{\rm th}({\epsilon}_{\gamma}>500{\rm GeV})=1.2\cdot 10^{-12}\cdot 
(\Theta
/0.1) {\rm cm}^{-2}{\rm sec}^{-1}$. Translating this to 3 TeV with a
spectral index ${\alpha}=1.1$ gives $F_{\gamma ,
\pi^0}^{\rm th}({\epsilon}_{\gamma}>3 {\rm TeV})=1.8\cdot
10^{-13}\cdot({\Theta}/0.1){\rm cm}^{-2}{\rm sec}^{-1}$, (Fig. 6). The
observed Cangaroo flux is $F_{\gamma}^{\rm 
obs}({\epsilon}_{\gamma}>3 {\rm
TeV})=3\cdot 10^{-12} {\rm cm}^{-2}{\rm sec}^{-1}\simeq 17 \cdot F_{\gamma
,\pi^0}^{\rm th}({\epsilon}_{\gamma}>3 {\rm TeV})$. Since in fact not 
only the
nonthermal X-rays but also the TeV \grs~ appear to come from only parts of
the SNR, one might increase the ratio of observed to predicted flux by
about 25 percent to a value of about 22. Even though SN 1006 should have
essentially reached the Sedov phase, we may nevertheless take
${\Theta}=0.2$ and ${\alpha}=1.0$, to
obtain (in this case) an upper bound of $F_{\gamma
,\pi^0}^{\rm th}({\epsilon}_{\gamma}>3 {\rm TeV})\simeq 7\cdot 10^{-13} {\rm
cm}^{-2}{\rm sec}^{-1}$, (Fig. 6). Reducing this flux by 25 percent, as
before, gives $F_{\gamma ,\pi^0}^{\rm th}({\epsilon}_{\gamma}>3 {\rm 
TeV})\simeq
5\cdot 10^{-13} {\rm cm}^{-2}{\rm sec}^{-1}\simeq (1/6)\cdot F_{\gamma}^{\rm
obs}({\epsilon}_{\gamma}>3{\rm TeV})$. Such an upper bound still
disregards the possibly higher gas compression ratio between the possibly
remaining ejecta and the shock which might exceed that assumed by DAV by a
factor up to $\sim 2$. Yet this additional factor is not certain and
should in any case not explain the above minimum discrepancy by a factor
of $\sim 6$. 

\noindent Thus, it does not seem possible to explain the observed flux by 
nucleonic
$\gamma$-rays without significantly revising the parameters $n,E_{SN},
{\rm and} \, d$, which are independent of acceleration theory.  Of course
such a revision can not be excluded, given the uncertainties. On 
the other
hand, the strong nonthermal X-ray emission is suggestive
of an IC explanation for SN 1006. This presupposes a fairly low magnetic
field, fitted to the radio and X-ray spectra as $B=3.5 {\eta}^{2/3}
\mu {\rm G}$ by Mastichiadis and de Jager (1996), for a gyro factor
$\eta \leq 30$.  Probably $1<{\eta}\ll10$ and then $3.5<B[{\mu}{\rm
G}]\ll 16$. Such low values in a SNR may be entirely possible if we
remember that the X-ray emission most likely comes from those regions
where the shock normal is roughly parallel to the external magnetic field
of several $\mu$G. Using the fit to the radio and X-ray spectra,
Mastichiadis and de Jager (1996) applied the simple electron acceleration
model of Mastichiadis (1996) to obtain an IC \gr flux that roughly agrees
with the observed Cangaroo flux provided we choose ${\eta} \simeq 3$, which 
is an acceptable value. 

\noindent Clearly, more detailed modelling is needed, but our preliminary 
conclusion is that SN~1006 is an electron IC source in TeV \grs. 

%subsection 8.2
\subsection{Tycho's SNR}
\noindent This 425 old SNR is probably approaching the end of the sweep-up 
phase (Tan and Gull, 1985). For an assumed external density $n\simeq 1{\rm
cm}^{-3}$, SN energy $E_{SN} \simeq 2\cdot 10^{50}{\rm erg}$ (Smith et al.,
1988), and distance $d\simeq 2.3$ kpc (Heavens, 1984), one obtains
$A\simeq 4\cdot 10^{-3}({\Theta}/0.1)$, similar to SN 1006. With $\alpha=1.1$
this gives $F_{\gamma ,{\pi}^0}^{\rm th}({\epsilon}_{\gamma}>500 {\rm 
GeV})=8\cdot 10^{-13}({\Theta}/0.1){\rm cm}^{-2}{\rm sec}^{-1}$ which for
${\epsilon}_{\gamma}> 300$ GeV lies a factor $\simeq 6\cdot ({\Theta}/0.1)$
below the upper limit of the Whipple observation (Lessard et al., 1997).
For $\alpha = 1.0$ and $\Theta = 0.2$, which is an entirely reasonable
choice for the evolutionary phase of Tycho, this factor reduces to $\simeq
1.8$, and if we would assume a postshock compression ratio $\sim$ twice
that of DAV, Whipple should just have missed a detection of Tycho ! (see Fig.
4)

\noindent With an angular diameter of 8 arc min, Tycho's SNR is 
essentially a 
point source for TeV \gr astronomy with consequent advantages for the
$\gamma$-hadron seperation. If the dominance of emission lines in the
X-ray spectrum ( Becker et al., 1980; Petre et al., 1993) can be taken as
an indication that the nonthermal electron component is very small, then a
detection of Tycho's SNR in TeV \grs~ may indeed be the first detection of
nuclear \grs~ in a CR source. And there may not be many more such sources
in the small sample of nearby candidates available. Thus every effort to
find TeV \grs~ in Tycho should be made. At the same time, it is clear that
only {\gr} {\it spectroscopy} together with detailed hard
X-ray continuum observations will ultimately enable us to clearly 
distinguish a $\pi^0$-decay source from an IC \gr source. 

%section 9
\section{Conclusions}
\noindent The conclusions from this discussion can be formulated quite 
concisely and we will simply do this here:
\begin{itemize}
\item SNRs fall into two major categories regarding TeV \gr emission
     \begin {itemize}
     \item SNRS in a uniform ISM and with progenitor masses $M \,< 20
     M_{\odot}$ are rapidly rising and very slowly decreasing TeV \gr 
     sources,
     except at very high ISM densities, where the upper cutoff energy may
     not reach present threshold energies of a few hundred GeV, either 
     because
     the SNR becomes too soon too weak to generate particles of sufficiently
     high energies, or because the ionisation in the upstream gas is too 
     low to allow strong wave excitation.
     \item SNRs in wind bubbles of massive
     progenitor stars $M \, > 20 M_{\odot}$ should have very low \gr emission
     (except at very early times) until reaching the swept-up shell of
     interstellar gas. The average magnetic field geometry is 
     unfavourable for efficient acceleration in the first place. 
     \end{itemize}
\item The theoretical estimates show that it is difficult but not
impossible to detect nearby SNRs in \grs.  These conclusions have been
dramatically confirmed by the recent detection of SN 1006. We 
interpret the present 
upper limits either as the result of too short observation times (Tycho) 
or of an unfavourable evolutionary phase in a massive star SNR.
\item The ambiguity of an electronic vs. a nuclear origin of the \gr emission
from SNRs can only be resolved by detailed spectrocopy, both in the
TeV and the hard X-ray regions. This observational effort needs to be
accompanied by detailed modelling of the synchrotron and IC emission
characteristics of the sources.
\item As in all other field of astronomy, detailed physics conclusions
are only possible in a {\it multi-wavelength approach}. For SNRs this
implies especially a reliable determination of such basic parameters
like distance, total explosion energy, and ambient density.
\end{itemize}

%section 10
\section{Acknowledgements}

\noindent I would like to thank Markus He\ss, Evgeny Berezhko, Michail 
Malkov and Felix Aharonian for
informative discussions on SNR physics, \gr emission, and acceleration
theory. 

\begin{refs}

\item Achterberg, A., Blandford, R.D., Reynolds, S.P. 1994, A\&A 281,220

\item Aharonian, F.A., Drury, L. O'C., V\"olk, H.J. 1994, A\&A 285, 645

\item Allen, G.E., et al. 1995, ApJ 448, L25

\item Allen, G.E., Geohane, J.W., Gotthelf, E.V., Petre, R., Jahoda, K., 
Rothschild, R.E., Lingenfelter, R.E., Heindl, W.A., Marsden, D., Gruber, D.E., 
Pelling, M.R., Blanco, P.R. 1997, to appear in ApJ Letters

\item Ammosov, A.E., Ksenofontov, L.T., Nikolaev, V.S., Pethukov, S.I. 1994, 
Astron. Lett. 20, 157

\item Asaoka, I., Aschenbach, B. 1994, A\&A 284, 573

\item Axford, W.I., Leer, E., Skadron, G. 1977, Proc. 15th ICRC (Plovdiv) 
11, 132

\item Axford, W.I., Leer, E., McKenzie, J.F. 1982, A\&A 111, 317

\item Baring, M.G., Ellison, D.C., Grenier, I. 1997a, in Proc. 2nd 
Integral Workshop, ESA, in press

\item Baring, M.G., Ellison, D.C., Reynolds, S.P., Grenier, I., Goret, P. 
1997b,(these proceedings)

\item Becker, R.H. et al. 1980, ApJ 235, L5

\item Bell, A.R. 1987, MNRAS 182, 443

\item Bennett, L., Ellison, D.C. 1995, J. Geophys. Res. 100, 3439             

\item Berezhko E.G. and Krymsky, G.F. 1988, Sov. Phys. Usp 12, 155

\item Berezhko, E.G., Yelshin, V.K., Ksenofontov, L.T. 1994, Astropart. 
Phys.2, 215

\item Berezhko, E.G., V\"olk, H.J. 1997a, Astropart. Phys. 7, 183

\item Berezhko, E.G., V\"olk, H.J. 1997b, in preparation

\item Berezinsky, V.S., Bulanov, S.V., Dogiel, V.A., Ginzburg, V.L., 
Ptuskin, V.S. 1990, {\it Astrophysics of Cosmic Rays}, North-Holland Publ. 
Comp., p. 18ff

\item Bhat C.L. et al. 1985, Nature 314, 511

\item Biermann, P.L. 1993a, A\&A 271, 649

\item Biermann, P.L. 1993b, Proc. 23rd ICRC (Calgary), Invited papers, 45

\item Blandford, R.D., Eichler, D. 1987, Phys. Rep. 154, 1

\item Bogdan, T. and V\"olk, H.J. 1983, A\&A 122, 129

\item Borione, A. et al. 1995, Proc. 24th ICRC (Rome) 2, 439

\item Brazier, K.T.S., Kanbach, G., Carraminana, A., Guichard, J., Merck, M. 
1996, MNRAS 281, 1033

\item Cesarsky, C.J. and Lagage, P.O. 1981, Proc 17th ICRC (Paris) 2, 335

\item Chevalier, R.A., Liang, E.P. 1989, ApJ 344, 332

\item Claussen, M.J., Frail, D.A., Goss, W.M., Gaume, R.A. 1997, to appear
in ApJ

\item Cong, H.I. 1977, Phd thesis, Columbia University

\item de Jager, O.C., Mastichiadis, A. 1997, ApJ 482, 874

\item Dorfi, E.A. 1990, A\&A 234, 419

\item Dorfi, E. 1991, A\&A 251, 597

\item Draine, B.T., McKee, C.F. 1993, ARAA 31, 373

\item Drury, L.O'C., V\"olk, H.J. 1981, ApJ 248, 344

\item Drury, L.O'C. 1983, Rep. Prog. Phys. 46, 973

\item Drury, L.O'C. Markiewicz, W.J., V\"{o}lk, H.J. 1989, A\&A 225, 179

\item Drury, L.O'C., Aharonian, F.A., V\"olk, H.J. 1994, A\&A 287, 959 (DAV)

\item Drury, L.O'C., Duffy, P., Kirk, J.G. 1996, A\&A 309, 1002

\item Eichler, D. 1984, ApJ 277, 429

\item Ellison, D.C., Eichler, D. 1984, ApJ 286, 691

\item Esposito, J.A., et al. 1996, ApJ 461, 820

\item Fesen, R.A. 1984, ApJ 281, 658  

\item Gaisser, T.K., Protheroe, R.J., Stanev, T. 1997, submitted to ApJ

\item Galeev, A.A. 1984, Sov. Phys. JETP 86, 1655

\item Galeev, A.A., Malkov. M.A., V\"olk, H.J. 1995, J. Plasma Physics 
part 1, 54, 59

\item Ginzburg, V.L. and Ptuskin, V.S. 1981, Proc. 17th ICRC (Paris) 2, 336

\item Heavens, A.F. 1984, MNRAS 211, 195

\item He\ss, M., for the HEGRA Collaboration, 1997, Proc. 25th ICRC
(Durban) 3, 229

\item Higdon, J.C., Lingenfelter, R.E. 1975, ApJ 198, L17

\item Jokipii, J.R. 1987, ApJ 313, 842

\item Jones, F.C. and Ellison, D.C. 1991, Space Sci. Rev. 58, 259

\item Jones, T.W., Kang, H. 1993, ApJ 402, 560

\item Kang, H., Jones, T.W. 1991, MNRAS 249, 439

\item Keohane, J.W., Petre, R., Gotthelf, E.V., Ozaki, M., Koyama, K. 1997, ApJ 
484, 350

\item Kirk, J.G., Duffy, P., Ball, L. 1995, A\&A 293, L37

\item Koyama, K., Petre, R., Gotthelf, E.V., Hwang, U., Matsuura, M., 
Ozaki, M., Holt, S.S. 1995, Nature 378, 255

\item Koyama, K., Kinugasa, K., Matsuzaki, K., Nishiuchi, M., Sugizaki, M., 
Torii, K., Yahauchi, S., Aschenbach, B. 1997, PASJ 49, in press

\item Lagage, P.O. and Cesarsky, C.J. 1983, A\&A 125, 249

\item Lebrun, F., Paul, J 1985, Proc. 19th ICRC (La Jolla) 1, 309

\item Lessard, R.W. et al. 1997, Proc. 25th ICRC (Durban) 3, 233

\item Levinson, A. 1994, ApJ 426, 327

\item Lisenfeld, U., Xu, C., V\"olk, H.J. 1996, A\&A 306, 677

\item Malkov, M.A., V\"olk, H.J. 1995, A\&A 300, 605

\item Markiewicz, W.J., Drury, L. O'C., V\"olk, H.J. 1990, A\&A 236, 487

\item Mastichiadis, A. 1996, A\&A 305, L53

\item Mastichiadis, A., de Jager, O.C. 1996, A\&A 311, L5

\item McClements, K.G., Dendy, R.O., Bingham, J., Kirk, J.G., Drury, L. 
O'C. 1997, MNRAS in press

\item McKenzie, J.F. and V\"olk, H.J. 1982, A\&A 116, 191

\item Moraal, H. and Axford, 1983, A\&A 125, 204

\item Mufson, S.L., McCollough, M.L., Dickel, J.R., Petre, R., White, R., 
Chevalier, R.A. 1986, Astron. J. 92, 1349

\item Naito, T., Takahara, J. 1994, J. Phys. G: Nucl. Part. Phys. 20, 477

\item Osborne, J.L., Wolfendale, A.W., Zhang, L. 1995, J. Phys. G: Nucl. 
Part. Phys. 21, 429

\item Petre, R. et al. 1993, in {\it UV and X-Ray Spectroscopy of Laboratory 
and Astrophysical Plasmas} (eds Silver, E. \& Kahn, S.), Cambridge Univ. 
Press, 424

\item Pollock, A.M.T. 1985, A\&A 150, 339

\item Prahl, J. and Prosch, C., for the HEGRA Collaboration, 1997 Proc 25th ICRC 
(Durban) 3, 217

\item Ptuskin, V.S., V\"{o}lk, H.J., Zirakashvili, V.N., Breitschwerdt, 
D. 1997, A\&A 321, 434

\item Reynolds, S.P. 1996, ApJ 459, L13

\item Reynolds, S.P. and Chevalier, R.A. 1981, ApJ 245, 912

\item Smith, A., Davelaar, J., Peacock, A. et al. 1988, ApJ 325, 288

\item Sturner, S.J., Skibo, J.G., Dermer, C.D. 1997, submitted to ApJ

\item Swordy, S.P., M\"{u}ller, D., Meyer, P. L'Heureux, J., Grunsfeld, 
J.M. 1990, ApJ 349, 625

\item Tan, S.M., Gull, S.F. 1985, MNRAS 216,949

\item The, L.-S., Leising, M.D., Kurfess, J.D., Johnson, W.N., Hartmann, D.H., 
Gehrels, N., Grove, J.E., Purcell, W.R. 1996, A\&AS 120, 357 

\item The, L.-S., Leising, M.D., Hartmann, D.H., Kurfess, J.D., Blanco, 
P., Bhattacharya, D. 1997, preprint astro-ph/9707086

\item V\"olk, H.J., Morfill, G.E., Forman, M.A. 1981, ApJ 249, 161

\item V\"olk, H.J., Zank, L., Zank, G. 1988, A\&A 188, 274

\end{refs}

\end{document}